\documentclass[twocolumn]{aa}
\usepackage{graphicx}
\begin{document}
   \title{Giant pulses in Pulsar \object{PSR B0031--07}}

   \author{A.~D. Kuzmin
          \inst{1}
          \inst{2}
          \and
          A.~A. Ershov
          \inst{1}
          \inst{2}
          }

   \offprints{A.~D. Kuzmin,
   \email{akuzmin@prao.psn.ru}}

   \institute{Pushchino Radio Astronomy Observatory, Astro Space
              Center, Lebedev Physical Institute,
              Russian Academy of Sciences, Pushchino, 142290, Russia
         \and Isaac Newton Institute, Chile, Pushchino Branch, Russia
             }

   \date{Received December 18, 2003; accepted July 20, 2004}

   \abstract{We report on  observations of the recently detected (Kuzmin et al. \cite{kuzmin04}) giant pulses
(GPs) from the pulsar PSR B0031--07 at 40 and 111 MHz. At 40 MHz
the peak flux density  of the strongest pulse is 1\,100 Jy, which
is 400 times as high as the peak flux density of the average pulse
(AP). A pulse whose observed peak flux exceeded the peak of the AP
by more than a factor of 200 is encountered approximately once in
800 observed periods. Peak flux density of the GPs compared to the
AP peak flux density $S^{\rm GP}_{\rm peak}/S^{\rm AP}_{\rm peak}$
has roughly a power-law distribution with a slope of $-4.5$. GPs
at 40 MHz are essentially stronger than those ones at 111 MHz.
This excess is approximately in inverse proportion to the
frequency ratio. The giant pulses are much narrower than the AP,
and cluster in two narrow regions of the AP near the peaks of the
two components of the AP. Some of the GPs emit at both phases and
are double. The separation of the double GP emission regions
depends on frequency. Similarly to the frequency dependence of the
width of the AP, it is less at 111 MHz than at 40 MHz. This
suggests that GPs are emitted from the same region of the
magnetosphere as the AP, that is in a hollow cone over the polar
cap instead of the light cylinder region. PSR B0031--07 as well as
the previously detected \object{PSR B1112+50} are the first
pulsars with GPs that do not have a high magnetic field at the
light cylinder. One may suggest that there are two classes of GPs,
one associated with high-energy emission from outer gaps, the
other associated with polar radio emission. The GPs of PSR
B0031--07 and PSR B1112+50 are of the second class. The dispersion
measure $DM$ is found to be
$10.900~\pm~0.003~\mathrm{pc~cm^{-3}}$.

    }

   \maketitle

\section{Introduction}

Giant pulses (GPs) are short-duration burst-like sudden increases
of intensity of individual radio pulses from pulsars. This rare
phenomenon has been detected previously only in five pulsars out of more than 1\,500 known ones:: the
\object{Crab pulsar} \object{PSR B0531+21} (Staelin \& Sutton
\cite{staelin}), PSR B1112+50 (Ershov \& Kuzmin \cite{ershov}),
the millisecond pulsars \object{PSR B1937+21} (Wolszczan et al.
\cite{wolszczan}) and \object{PSR B1821--24} (Romani \& Johnston
\cite{romani}), and the extragalactic pulsar \object{PSR
B0540--69} (Johnston \& Romani \cite{johnston}).

The intensities of the GPs are extremely high. The brightness
temperature of GPs of the millisecond pulsar PSR B1937+21 is as
high as $T_{\rm B} \approx 10^{35}$~K (Kuzmin \& Losovsky
\cite{kuzmin02}), and  $T_{\rm B} \ge 10^{36}$~K for GPs of the
Crab pulsar (Kostyuk et al. \cite{kostyuk}, Hankins et al.
\cite{hankins}).

Four of these pulsars belong to the group of pulsars with the
strongest magnetic field at the light cylinder. Pulsar PSR
B1112+50 is the first exception from this group, being a pulsar that
exhibits the characteristic features of GPs but has an ordinary
value of the magnetic field on the light cylinder. Kuzmin et
al.(\cite{kuzmin04}) recently reported the detection of giant pulses
from another ordinary magnetic field pulsar, PSR B0031--07. They
find that at 111 MHz the peak flux density of the strongest GP is
530 Jy, which is 120 times as high as than the peak flux density of
the average profile (AP). A pulse whose observed peak fluxes
exceed the peak of the AP by more than a factor of 50 is
encountered approximately once in 250 observed periods. Giant
pulses cluster in a narrow region near  the peak of the first
component of the AP.

In this paper we report observations of GPs of pulsar PSR
B0031--07 at the second frequency of 40 MHz, additional
observations at 111 MHz and a comparative analysis of the
two-frequency data.

\section{Observations and data reduction}

Observations were carried out at two frequencies, 40 and 111 MHz.

At 40 MHz observations were performed from August 28 through
October 13, 2003 with the DKR Radio Telescope at Pushchino Radio
Astronomy Observatory of the Lebedev Physical Institute. The DKR
telescope has an effective area of about 8\,000 square meters. One
linear polarization was received. We used a 128-channel receiver
with channel bandwidth 1.25 kHz. The frequency of the first
(highest frequency) channel was 40.982 MHz, the sampling interval
was 0.819 ms, and the time constant was $\tau = 1$~ms. The
duration of each observation was about 16 min (994 pulsar
periods). A total of 16 observations containing 15\,904 pulsar
periods was carried out.

From June 21 through September 22, 2003 we performed an additional
16  observations at 111 MHz and 12 simultaneous observations at
111 and 40 MHz. At 111 MHz we used the Large Phased Array (BSA)
Radio Telescope at Pushchino Radio Astronomy Observatory with an
effective area of about 20\,000 square meters. One linear
polarization was received. At both frequencies we used two
64-channel receivers with channel bandwidth 20 kHz. The sampling
interval was 2.560 ms and the time constant was $\tau = 3$~ms. The
durations of the observations were about 3 min (205 pulsar
periods) and 9 min (600 pulsar periods) at 111 and 40 MHz
respectively.

All observations were time-referenced to the Observatory rubidium
master clock, which in turn was monitored against the National
Time standard via TV timing signals. Pulsar ephemerides were
obtained from the catalogue by Taylor et al. (\cite{taylor}).

During the off-line data reduction the signal records were cleaned
of radio interferences. Subsequently the inter-channel
dispersion delays imposed by an interstellar medium were removed.
Each observing period was analyzed for pulses with amplitude
exceeding a preset level and its amplitude, pulse width and phase
were derived.

\section{Results}

At 40 MHz the 212 pulses whose observed peak fluxes density
exceeded the peak flux density of the average profile (AP) (which
is equivalent to the average pulse) by more than a factor of 100
were selected and analyzed. 19 of them exceeded the peak of the AP by
more than a factor of 200. A pulse whose observed peak fluxes
exceeded the peak of the AP by more than a factor of 100 is
encountered approximately once in 75 observed periods.

   \begin{figure}
     \centering
      \resizebox{\hsize}{!}{\includegraphics{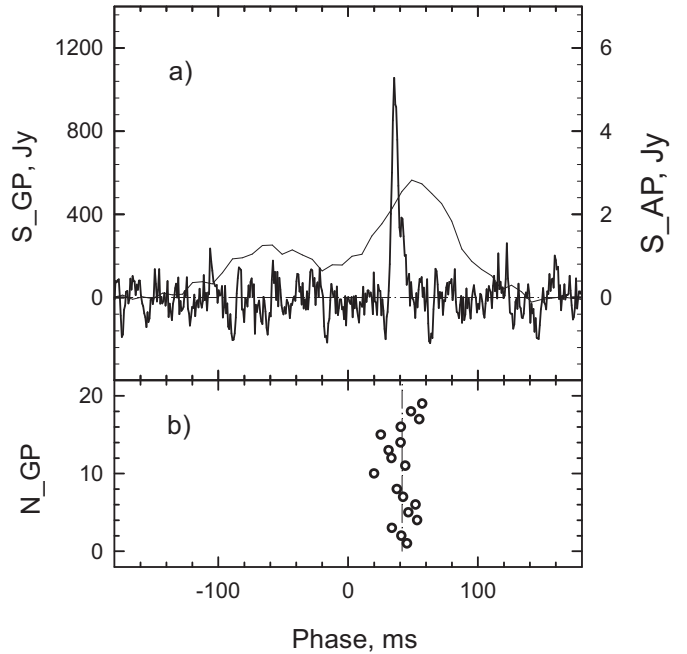}}
      \caption{\textbf{a)}
       The strongest observed GP at 40 MHz (bold line)
       together with the AP (thin line). The plot of the AP is presented on a
       200 times larger scale and flux densities of the observed GP and AP are shown
       separately on the left and right sides of the "y" scales. \textbf{b)} The
       phases of the observed GPs relative to the center of the AP
       determined as the mid-point between 0.1 intensity levels.
              }
         \label{Fig1}
   \end{figure}

Figure~\ref{Fig1}a shows the strongest observed GP (bold line)
together with the AP for all 16 days of observations. The observed
peak flux density of this GP exceeds the peak flux density of the
AP by factor of 400. The plot of the AP is presented on a 200
times larger scale and flux densities of the observed GP and AP
are shown separately on the left and right sides of the "y"
scales.

The value of the GP flux density was determined relative to the
known AP flux density as

$$
  S_{\rm peak}^{\rm GP} = S_{\rm peak}^{\rm AP} \times (I^{\rm GP} /I^{\rm AP})~,
$$

where $I^{\rm GP} /I^{\rm AP}$ is the ratio of the GP to AP peak
intensity and $S_{\rm peak}^{\rm AP}$ is the peak flux density of
AP

$$
  S_{\rm peak}^{\rm AP} = S_{\rm mean}^{\rm AP} /k_{\rm form}~,
$$

where $S_{\rm mean}^{\rm AP}$ is the flux density averaged over a
pulsar period, $k_{\rm form}$ is the duty cycle of the pulsar.

The value of the AP flux density $S_{\rm mean}^{\rm AP} = 490$~mJy
was obtained from Izvekova et al. (\cite{izvekova81}). The reduction
factor is $k_{\rm form} = 0.18,$ so the AP peak flux density is
2.7 Jy. The strongest observed GP has $I^{\rm GP} /I^{\rm AP} =
400$. Then, the observed GP peak flux density is $S_{\rm peak}^{\rm
GP} = 2.7 \times 400 = 1\,100$ Jy.  The intrinsic fine structure
of the pulses is masked by interstellar scattering. In our
measurements of the 19 strongest GPs at 40.9 MHz we derived $\tau_{\rm
sc} = 3.1~\pm~0.4$ ms. Scattering increases the pulse width and
decreases the pulse amplitude. Thus the intrinsic GP peak flux
density is $S_{\rm peak}^{\rm GP} > 1\,100$ Jy.

   \begin{figure}
      \centering
      \resizebox{\hsize}{!}{\includegraphics{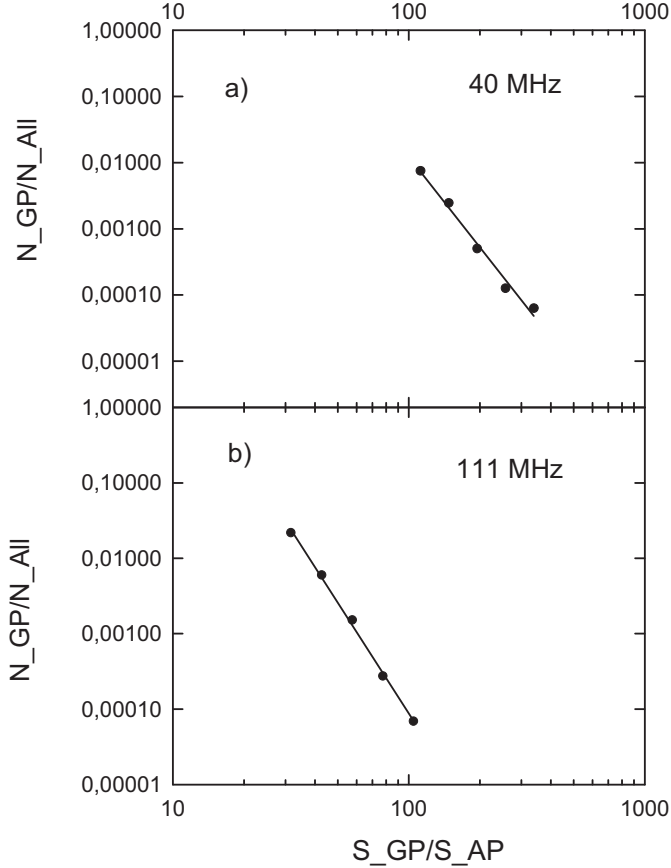}}
      \caption{\textbf{a)}
       The cumulative distribution of the  observed GP peak flux
       density as related to the AP peak flux density for 40 MHz,
       \textbf{b)} The same for 111 MHz.
              }
         \label{Fig2}
   \end{figure}

In Fig.~\ref{Fig2}a we show the cumulative distribution of the
ratio of the observed GP peak flux density to the AP peak flux
density $S^{\rm GP}_{\rm peak}/S^{\rm AP}_{\rm peak}$ of all
pulses that we have selected and analyzed. In a detectable
interval of $S^{\rm GP}_{\rm peak}/S^{\rm AP}_{\rm peak}$ from 100
to 400, the histogram has roughly a power-law distribution with a
slope of $-4.5$. Since for $S^{\rm GP}_{\rm peak}/S^{\rm AP}_{\rm
peak} < 100$ the pulse signal-to-noise ratio is less than 5, at
lower intensities the observed distribution is masked by noise.

The distribution for 111 MHz, which we have obtained from the Kuzmin et
al. (\cite{kuzmin04}) data and additional observations, is shown in
Fig.~\ref{Fig2}b. In a detectable interval of $S^{\rm GP}_{\rm
peak}/S^{\rm AP}_{\rm peak}$ from 30 to 100, the histogram has
roughly a power-law distribution with a slope of $-4.8$. Since for
$S^{\rm GP}_{\rm peak}/S^{\rm AP}_{\rm peak} < 20$ the pulse
signal-to-noise ratio is less than 5, at lower intensities the
observed distribution is masked by noise.

The distributions in Fig.~\ref{Fig2} demonstrate, that GPs at 40
MHz are essentially stronger than those at 111 MHz. At 40 MHz one
in every thousand pulses exceeds the AP by more than a factor of 170,
but at 111 MHz only by 60. The strongest observed GP at 40 MHz
exceeds the peak flux density of the AP by a factor of 400, but
at 111 MHz only by a factor of 120. In a quantitative sense this
gain is approximately in inverse proportion to the ratio of
frequencies.

From Fig.~\ref{Fig2} one can see that the rate of GPs depends on
the frequency. At 40 MHz a pulse whose observed peak flux
exceeds the peak of AP by more than a factor 100 is encountered
once in 85 observed periods, but at 111 MHz only once in 11\,000.
This may be a reason why there are no claims of GPs from this
pulsar at higher frequencies.

Our distributions cover a smaller number of observed pulsar
periods (by two orders of magnitude) compared to such histograms
for previously known short period pulsars. Therefore, we consider
the slope index as a tentative one which needs to be refined.

The mean observed width\footnote{The width is not corrected for
scatter and dispersion broadening} of 19 GPs with $S^{\rm GP}_{\rm
peak}/S^{\rm AP}_{\rm peak} > 200$ is $w_{\rm 50}^{\rm obs} = 6~\pm~2$ ms.
The width  of the AP\footnote{The width of the AP was
measured at half the maximum height of the leading and trailing
components.} for all 12 days of observations is $w_{\rm 50}^{\rm AP} =
180$ ms.  Thus, at 40 MHz GPs are narrower than the AP by about a
factor of 30.

The positions of the GPs are stable within the AP and they cluster
in a narrow window near the maximum of the second component of the
AP. Figure~\ref{Fig1}b shows the phases $\Phi$ of the GPs which
exceeded the AP by factor of 100.  The average phase difference
between the GPs and the center of the AP is $\Phi_{\rm 40} \cong 40$
ms. GPs cluster in a narrow phase window $\Delta \Phi = \pm 10$
ms. The clustering is tighter for stronger GPs.

The GPs are not identified with subpulses,  since contrary to the
stable phase position of GPs, the phase position of subpulses
drift across the  AP. GPs are also narrower than the subpulses of
this pulsar $w_{\rm 50}=15$ ms at 60 and 102 MHz (Izvekova et al.
\cite{izvekova93}).

The GP brightness temperature is

$$
  T_{\rm B} = S \lambda^2 /2k\Omega~.
$$

Here $\lambda$ is the radio wavelength, $k$ is the Boltzmann
constant, and $\Omega$ is the solid angle of the radio emission
region. Adopting $\Omega \simeq (l/d)^2$, where $l$ is the size of
the radio emission region and $d$ the distance to the pulsar, and further
adopting $l \leq  c \times w_{\rm 50}$, where $c$ is the speed of
light and the distance to the pulsar $d = 0.68$ kpc (Taylor et al.
\cite{taylor}), we obtain $T_{\rm B} \geq 10^{28}$ K.

   \begin{figure}
      \centering
      \resizebox{\hsize}{!}{\includegraphics{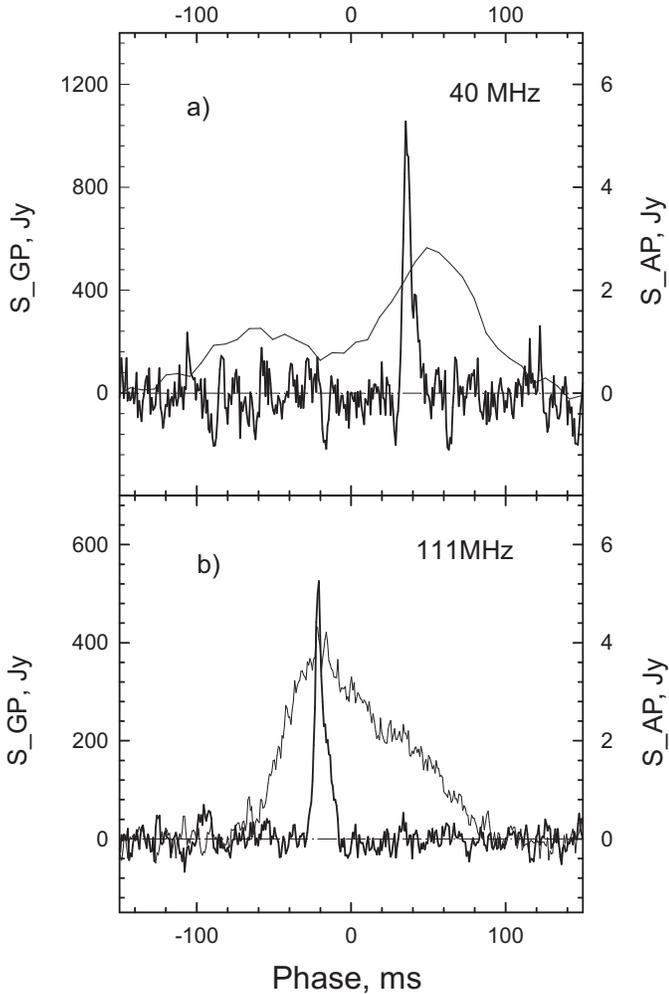}}
      \caption{ \textbf{a)} The strongest observed GP at 40 MHz (bold
      line),
       together with the AP (thin line), \textbf{b)} The strongest observed
       GP at 111 MHz (bold line), together with the AP (thin line).
       Plots of GPs and APs are presented on different scales,
       shown on the left and right sides of the "y" axis.
              }
         \label{Fig3}
   \end{figure}

The most unexpected and interesting feature of our observations at
40 MHz is the non-coincidence of the phase positions of the GPs at 40 and
111 MHz. In Fig.~\ref{Fig3} we present the time-alignment of GPs
and APs at 40 and 111 MHz (Kuzmin et al. \cite{kuzmin04}). One can
see that the GP at 40 MHz is separated from the GP at 111 MHz by
about 60 ms and that they are located in two different parts of the AP. At 111
MHz the GP is located near the maximum of the first component of
the AP, whereas at 40 MHz the GP is located near the maximum of
the second component.

To verify this peculiarity, we have performed a control
determination of the dispersion measure $DM$. For this we varied
the value of $DM$ by a process of removal of the inter-channel
dispersion delays and searched their optimal value, which provides
the maximum amplitude of individual pulses. We processed  440
individual pulses at 111 MHz and 337 at 40 MHz. The mean
value of dispersion measure was obtained to be $DM = 10.900~\pm~0.003~\mathrm{pc~cm^{-3}}$.
This value is consistent with the
catalog value $10.89~\pm~0.01~\mathrm{pc~cm^{-3}}$ (Taylor et
al. \cite{taylor}) and confirms the validity of the non-coincidence
of the phase positions of the GPs at 40 and 111 MHz.

One may suggest two explanations of such behavior. The first is
that the emission region of GPs moves with frequency. The second
proposes that there are two separate phase emission regions
for the GPs. In this case, one may expect that some of the GPs
will emit at both phases and are double.

To clarify this point, we undertook a more detailed inspection of
the GPs. Our first-line reduction program searched only for a single GP
per pulsar period, whose amplitude exceeded a preset level.
However, one may expect that, in the case of two emission regions,
some GPs have two components, where the first component is
observed mainly at 111 MHz and the second
at 40 MHz. Therefore we performed an addition search
of the observed GPs for two-component structure. For this search
we used a 20 kHz channel bandwidth at both frequencies. The answer
is yes: --- some GPs, both at 40 and 111 MHz, have a two-component
structure. We detected 12 double GPs whose amplitude
exceeded the AP by a factor of 50  at 111 MHz  and 5 double GPs with
amplitude exceeding AP by a factor of 20 at 40 MHz.

   \begin{figure}
      \centering
      \resizebox{\hsize}{!}{\includegraphics{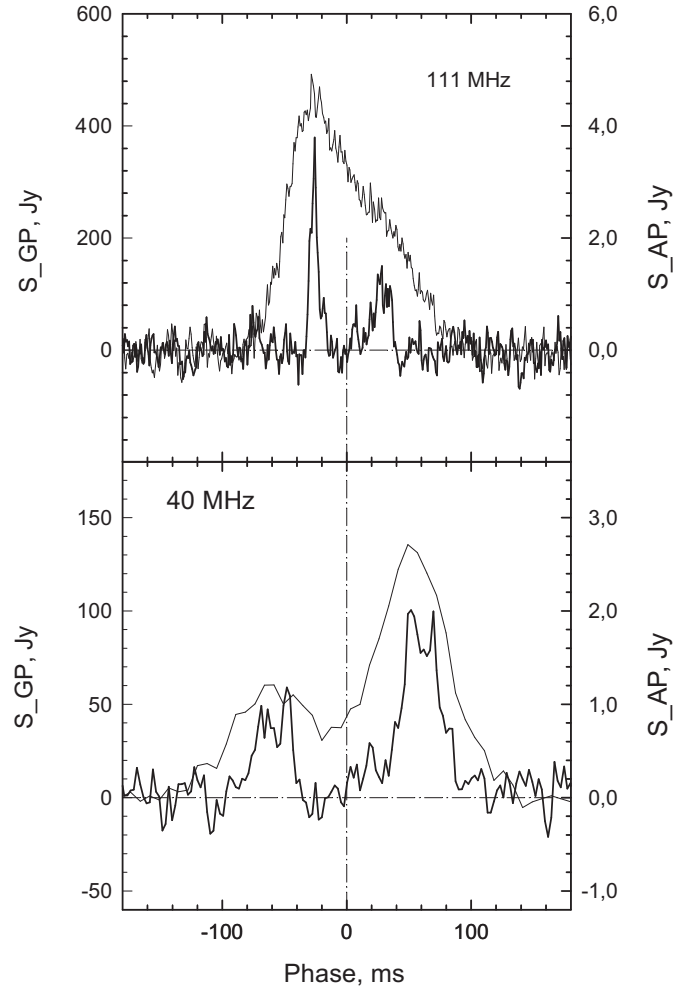}}
      \caption{ \textbf{a)} The double GP (bold line) observed
      at 111 MHz together with the AP (thin line),
         \textbf{b)} The double GP (bold line), observed at 40 MHz.
         together with the AP (thin line).
              }
         \label{Fig4}
   \end{figure}

An example of two-component GPs is shown in Fig.~\ref{Fig4}.
Since at 40 MHz double pulses were searched in a 20 kHz channel
bandwidth, the dispersion pulse broadening of 30 ms
reduces the strength of the GP and distorts the observed
pulse shape compared to other GPs, which were observed in 1.25
kHz channel bandwidth.

   \begin{figure}
      \centering
      \resizebox{\hsize}{!}{\includegraphics{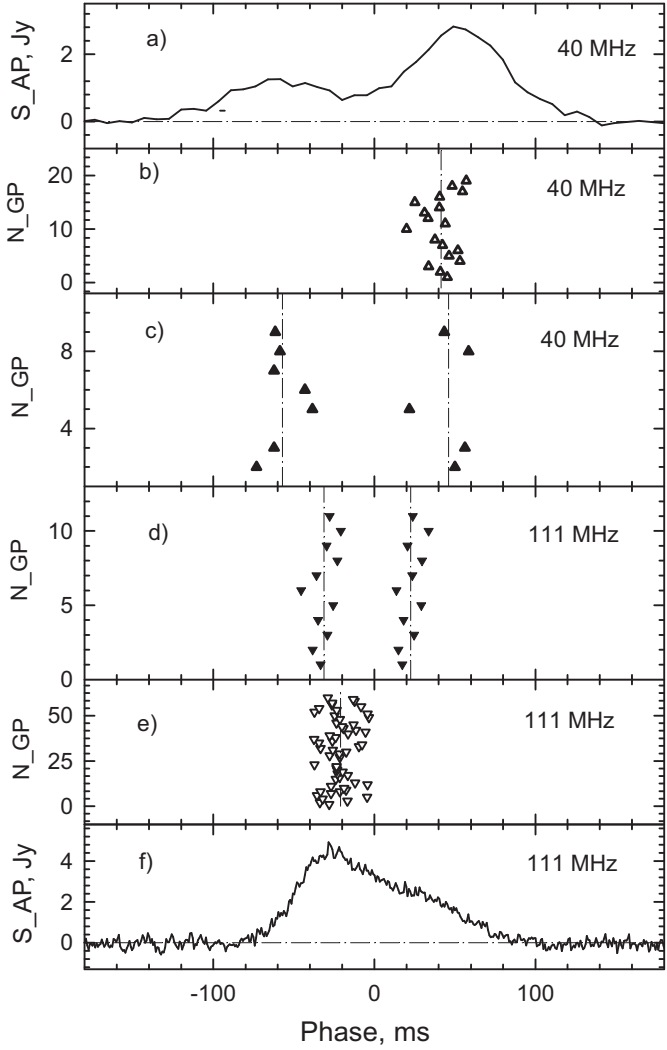}}
      \caption{ \textbf{a)} Integrated profile at 40 MHz,
      \textbf{b)} Phases of the observed single GPs at 40 MHz,
       \textbf{c)} Phases of the observed double GPs at 40 MHz,
        \textbf{d)} Phases of the observed double GPs at 111 MHz,
         \textbf{e)} Phases of the observed single GPs at 111 MHz,
          \textbf{f)} Integrated profile at 111 MHz.
              }
         \label{Fig5}
   \end{figure}

The phases of these double GPs, as well as single GPs from our
observations and the Kuzmin et al. (\cite{kuzmin04}) data are plotted in
Fig.~\ref{Fig5} together with APs.

One can see that the phases of the trailing component of double
GPs at 40 MHz (Fig.~\ref{Fig5}c) fit the phases of the single GPs at this
frequency (Fig.~\ref{Fig5}b). In a similar manner, the phases of the leading
component of double GPs at 111 MHz (Fig.~\ref{Fig5}d) fit the phases of the single
GPs at 111 MHz (Fig.~\ref{Fig5}e). This supports the concept of two emission
regions of GPs. The first dominates at 111 MHz, the second
dominates at 40 MHz and both form double GPs.

Twelve observation sessions,  containing 2\,460 pulsar periods,
were performed simultaneously at 40 and 111 MHz. We also used the 20
kHz channel bandwidth at both frequencies. Forty pulses at 40 MHz
and 45 pulses at 111 MHz with peak intensities exceeding the peak
intensity of the AP by more than a factor of 20 were selected and
analyzed. Only two among these pulses were found in the same pulse
phase inside the same pulsar period and identified as
simultaneous. The probability that there are  accidental
coincidences of GPs simultaneously at two frequencies is equal to
$P_{\rm 40}*P_{\rm 111}=40/2460*45/2460 \approx 10^{-7}$; this is
significantly lower than the observed detection rate of
simultaneous GPs of $2/2460 \approx 10^{-3}$.

The measured spectral indices between frequencies of 40 and 111 MHz
for these 2 simultaneous GPs are $\alpha^{\rm GP}=-0.6$ and
$-1.1$, corresponding to the low-frequency turn-over  region of
this pulsar (the spectral index of the AP is $\alpha^{\rm
AP}=-0.2$). However, only a small number of simultaneous  GPs was
observed. This possibility therefore needs further study.

\section{Discussion}

The GPs that we detected from PSR B0031--07 exhibit all
characteristic features of the classical GPs from PSR B0531+21 and
PSR B1937+21.

The peak intensities of the GPs exceed the peak intensity of the AP by
more than a factor of 400. The histograms of the flux density have
a power-law distribution. The GPs are much narrower than the
AP and their phases are stable inside the integrated profile.

Alongside these similarities, the GPs of the pulsar PSR B0031--07 have
noticeable differences. Previously known pulsars with GPs belong
to the group of pulsars with very high magnetic field on the light
cylinder. Pulsar PSR B0031--07, as well as the previously detected
PSR B1112+50 (Ershov \& Kuzmin \cite{ershov}), are the first ones
with ordinary magnetic fields on the light cylinder.

The most unexpected and interesting result is the clustering of
the GPs in two different regions. This indicates that there are
two emission regions of GPs. The separation of these regions at 40
MHz is larger than at 111 MHz; this is corresponds to the increase in
the width of the AP, which is interpreted as a divergence of the
magnetic field lines in the hollow cone model of pulsar radio
emission.  This suggests that the GPs from this pulsar, and
possibly from PSR B1112+50, originate in the same region as the
AP, that is in a hollow cone over the polar cap instead of in the
light cylinder region.

One may suggest that there are two classes of GPs, one associated
with high-energy emission from the outer gaps, the other
associated with polar radio emission. The GPs of PSR B0031--07 and
PSR B1112+50 are of the second class.

\section{Conclusions}

The detection of giant pulses from pulsar PSR B0031--07 has been
confirmed.  The peak flux density of the strongest pulse is 1\,100
Jy, which is 400 times as high as than the peak flux density of the
average profile.

The GPs are much narrower than the average profile and cluster in
two narrow regions of the average profile. The separation of the
two GP emission regions decreases with frequency. This suggests
that the GPs from this pulsar are emitted from the same region as
the AP, that is in the hollow polar cone instead of in the light
cylinder region.

Pulsar PSR B0031--07, as well as the previously detected PSR
B1112+50, are the first pulsars with GPs that do not have a high
magnetic field at the light cylinder, such as previously known pulsars
with GPs have. They may be a separate class of GPs.

The dispersion measure $DM$ is found to be 10.900 $\pm$ 0.003
$\mathrm{pc~cm^{-3}}$.

 \begin{acknowledgements}

We wish to thank  V.~V. Ivanova, K.~A. Lapaev \& A.~S. Aleksandrov
for assistance during observations. We are grateful to Francis
Graham Smith for improving the English and valuable comments.
We are grateful to the anonymous referee  for comments which
enhanced the paper. This work was supported in part by the
Russian Foundation for Basic Research (project No 01-02-16326) and
the Program of the Presidium of the Russian Academy of Sciences
"Non-steady-state Processes in Astronomy."

\end{acknowledgements}



\begin{thebibliography}{}

   \bibitem[2003]{ershov}
     Ershov, A.~A. \& Kuzmin, A.~D. 2003, Pis'ma v AZh, 29, 111 (Astr. Lett., 29, 91)

   \bibitem[2003]{hankins}
     Hankins, T.~H., Kern, J.~S., Weatherall, J.~C., \& Eilek,
     J.~A. 2003, Nature, 422, 141

   \bibitem[1981]{izvekova81}
     Izvekova, V.~A., Kuzmin, A.~D., Malofeev, V.~M., \& Shitov, Yu.~P. 1981,
     Ap\&SS, 78, 45

   \bibitem[1993]{izvekova93}
     Izvekova, V.~A., Kuzmin, A.~D., Lyne, A.~G., et al. 1993,
     MNRAS, 261, 865

   \bibitem[2003]{johnston}
     Johnston, S., \& Romani, R.~W. 2003, ApJ, 590, L95

   \bibitem[2003]{kostyuk}
     Kostyuk, S.~V., Kondratiev, V.~I., Kuzmin, A.~D., Popov, M.~V.,
     \& Soglasnov, V.~A. 2003, Pis'ma v AZh, 29, 440 (Astr. Lett., 29,
     387)

   \bibitem[2002]{kuzmin02}
     Kuzmin, A.~D., \& Losovsky, B.~Ya. 2002, Pis'ma v AZh, 28, 25 (Astr. Lett., 28,
     21)

   \bibitem[2004]{kuzmin04}
     Kuzmin, A.~D., Ershov, A.~A., \& Losovsky, B.~Ya. 2004, Pis'ma v
     AZh, 30, 285 (Astr. Lett., 30, 247)

   \bibitem[2001]{romani}
     Romani, R.~W., \& Johnston, S. 2001, ApJ, 557, L97

   \bibitem[1970]{staelin}
     Staelin, D.~H., \& Sutton, J.~M. 1970, Nature, 226, 69

   \bibitem[1995]{taylor}
     Taylor, J.~H., Manchester, R.~N., Lyne, A.~G., et al. 1995,
     Catalog of 706 Pulsars, unpublished work


  \bibitem[1984]{wolszczan}
    Wolszczan, A., Cordes, J.~M., \& Stinebring, D.~R. 1984,
    in Millisecond Pulsars,
    ed.\ S.~P. Reynolds \& D.~R. Stinebring
    (NRAO, Green Bank), p.63

\end{thebibliography}
\end{document}